\documentclass[a4paper,fleqn,usenatbib]{mnras}
\usepackage{mathptmx}

\usepackage[T1]{fontenc}
\usepackage{ae,aecompl}

\usepackage{graphicx}	
\usepackage{amsmath}	
\usepackage{amssymb}	
\usepackage{dcolumn}
\usepackage{epstopdf}
\usepackage{hyperref}

\newcommand{\cms}[1]{cm$^{3}$.s$^{-1}$#1} 

\title[Quantum mechanical study of the high-temperature $\mathrm{H}^+ + \mathrm{HD} \to \mathrm{D}^+ + \mathrm{H}_2$ reaction...]
      {Quantum mechanical study of the high-temperature $\mathrm{H}^+ + \mathrm{HD} \to \mathrm{D}^+ + \mathrm{H}_2$ reaction for the primordial universe chemistry}

\author[M. Lepers et al.]{
Maxence Lepers,\thanks{E-mail: maxence.lepers@u-bourgogne.fr}
Gr{\'e}goire Guillon,
Pascal Honvault
\\
Laboratoire Interdisciplinaire Carnot de Bourgogne, CNRS, Universit\'e de Bourgogne Franche-Comt\'e, 21078 Dijon, France
}

\date{Accepted XXX. Received YYY; in original form ZZZ}

\pubyear{2019}

\begin{document}
\label{firstpage}
\pagerange{\pageref{firstpage}--\pageref{lastpage}}
\maketitle

\begin{abstract}
We use the time-independent quantum-mechanical formulation of reactive collisions in order to investigate the state-to-state $\mathrm{H}^+ + \mathrm{HD} \to \mathrm{D}^+ + \mathrm{H}_2$ chemical reaction. We compute cross sections for collision energies up to 1.8 electron-volts and rate coefficients for temperatures up to 10000 kelvin. We consider HD in the lowest vibrational level $v=0$ and rotational levels $j=0$ to 4, and H$_2$ in vibrational levels $v'=0$ to 3 and rotational levels $j'=0$ to 9. For temperatures below 4000 kelvin, the rate coefficients strongly vary with the initial rotational level $j$, depending on whether the reaction is endothermic ($j\le 2$) or exothermic ($j\ge 3$). The reaction is also found less and less probable as the final vibrational quantum number $v'$ increases. Our results illustrate the importance of studying state-to-state reactions, in the context of the chemistry of the primordial universe.
\end{abstract}




\section{Introduction}

It is now well established that the molecules composing the primordial universe, that contained neutral or singly-ionised hydrogen (H), deuterium (D), helium ($^4$He) and lithium ($^7$Li), were out of thermodynamic equilibrium \citep{coppola2011, coppola2013, coppola2016}. In other words, the populations in their rovibrational levels did not follow a Maxwell-Boltzman distribution. Therefore, in order to understand the physical and chemical evolution of the primordial gas, which led in particular to the formation of the first stars, it is necessary to have quantitative information about the chemical reactions involving different rovibrational levels of the reactants and the products, so-called \textit{state-to-state} reactions \citep{galli2013, bovino2018}. In particular, rate coefficients are needed on a wide range of temperatures, up to a few thousand kelvin (K). Theoretical calculations of rate coefficients are welcome, since they allow for covering such a wide range, in a more accessible way 	than experimental measurements.

In this context, the molecules H$_2$ and HD play a central role, as they act as coolants of the primordial gas, through the mechanism of collisional excitation to higher rovibrational levels followed by the spontaneous emission of a photon. This cooling process has a strong influence on the gravitational collapse leading to the first structures. Below 500~K, HD is the main coolant, due to its electric dipole moment (about $10^{-3}$ debye) and its smaller rovibrational spacings compared to H$_2$ (see for instance \citet{galli2002, ripamonti2007, kreckel2010}). Therefore, it is crucial to investigate the chemical reactions involving HD, especially with neutral and ionised atoms, in order to characterise the cooling dynamics. If the reaction with the most abundant species H has been widely studied (see for instance \citep{flower1999, ely2016, desrousseaux2018}), very little is known about its ionic counterpart $\mathrm{H}^+ + \mathrm{HD}$, which is the subject of the present article.

In the literature, the scarce results given on the $\mathrm{H}^+ + \mathrm{HD} \to \mathrm{D}^+ + \mathrm{H}_2$ reaction \citep{henchman1981, smith1982, millar1989, gerlich2002, jambrina2012} generally belong to articles focused on the inverse reaction $\mathrm{D}^+ + \mathrm{H}_2 \to \mathrm{H}^+ + \mathrm{HD}$ \citep{fehsenfeld1974, gerlich1992, jambrina2009, honvault2013, gonzalez-lezana2013, sahoo2014, sahoo2015, lara2015, bhowmick2018}. In Refs~\citep{henchman1981, smith1982}, thermal rates of $1.1 \pm 0.2$ and $1.7\pm 0.2 \times 10^{-10}$~\cms~are measured at 205 and 295~K respectively using the selected ion flow tube (SIFT) technique. \citet{millar1989} derived an Arrhenius-type formula from those measurements. Later, \citet{gerlich2002} also gave an Arrhenius-type formula fitted from most dynamically biased (MDB) statistical calculations between 30 and 130~K. Finally, \citet{jambrina2012} give thermal rate coefficients calculated both with a time-independent quantum-mechanical (TIQM) method and with variants of the quasi-classical trajectory (QCT) method on the so-called ARTSM potential-energy surface (PES) by \citet{aguado2000}. The agreement of the TIQM results with the above-mentioned experiments is good, even if the latter are better reproduced by the statistical QCT calculations at 295~K. It is worthwhile noting that the reactive collision between Rydberg hydrogen atoms and HD has been investigated experimentally \citep{yu2014}.

The weak interest to the $\mathrm{H}^+ + \mathrm{HD} \to \mathrm{D}^+ + \mathrm{H}_2$ reaction is probably due to its endothermiciity (for HD and H$_2$ in their ground rovibrational levels $(v,j)=(0,0)$ and $(v',j')=(0,0)$). This endoothermicity, equal to 39.5 meV, mainly comes from the difference in zero-point energies between H$_2$ and HD (35.8~meV), but also from the difference in ionisation potentials of H and D (3.7~meV). But firstly, this activation energy is widely overcome at the temperatures that we consider here; and secondly, the title reaction becomes exothermic for HD in rovibrational levels higher than $(v=0,j=3)$. On the other hand, for electronic energies larger than 1.83~eV above the dissociation limit $\mathrm{H}^+ + \mathrm{H}_2(X^1\Sigma_g^+)$, charge transfer becomes possible towards the channel $\mathrm{H}(^2S) + \mathrm{H}_2^+(X^2\Sigma_g^+)$. However, time-dependent wave-packet (TDWP) calculations for $\mathrm{D}^+ + \mathrm{H}_2$ \citep{ghosh2015} and $\mathrm{H}^+ + \mathrm{H}_2$ \citep{ghosh2017} on the three lowest diabatic PESs of H$_3^+$ \citep{viegas2007} have shown that the charge transfer processes are much less probable than the reactive one without charge transfer.

In this article, we compute the cross sections and the rate coefficients characterizing the reaction $\mathrm{H}^+ + \mathrm{HD}(v,j) \to \mathrm{D}^+ + \mathrm{H}_2(v',j')$, with HD in the lowest vibrational level $v=0$ and rotational levels $j=0$ to 4, and H$_2$ in vibrational levels $v'=0$ to 3 and rotational levels $j'=0$ to 9. Our calculations are performed with the TIQM method for reactive collisions, based on hyperspherical coordinates, which take into account the indistinguishability of the two H nuclei. Within the framework of the Born-Oppenheimer approximation, we characterize the motion of the two H and the D nuclei on the so-called VLABP ground state global PES of H$_3^+$ calculated by \citet{velilla2008}, an improved version of the ARTSM PES \citep{aguado2000} that cautiously takes into account long-range interactions. Because we ignore the hyperfine interactions, we do not account for the difference in ionisation potentials between H and D, which introduces an uncertainty of 3~meV on collisional energy. But in this study, we are more interested in large collision energies, up to 1.8~eV above the lowest channel $\mathrm{H}^+ + \mathrm{HD}(0,0)$, for which we can ignore charge transfer. Because our cross sections are very low at 1.8~eV, we can give converged rate coefficients for temperatures up to 10000~K for the lowest rovibrational levels of H$_2$.
We also calculate thermal rate coefficients, assuming HD rovibrational levels in thermodynamic equilibrium, in order to compare our results with literature, and find larger rates. We possibly attribute the discrepancies with previous TIQM results to differences in the asymptotic region of the underlying PESs. Regarding previous experimental results, rate coefficients are given at only two temperatures, and so we think that additional measurements would be particularly relevant.

This article is organised as follows. Section \ref{sec:meth} describes our TIQM method, giving in particular the expressions of the cross section and rate coefficient. Section \ref{sec:res} presents our results, dealing with HD in the rovibrational ground level (Sec.~\ref{sub:j0}), in rotationally-excited levels (Sec.~\ref{sub:j1}), and in thermodynamic equilibrium (Sec.~\ref{sub:therm-eq}), to allow comparison with literature results. Section \ref{sec:ccl} contains concluding remarks.

\section{Method}
\label{sec:meth}

In this article, we focus on two quantities characterizing the reactive collision $\mathrm{H}^+ + \mathrm{HD}(v,j) \to \mathrm{D}^+ + \mathrm{H}_2(v',j')$: the state-to-state cross section $\sigma_{vj,v'j'}$ given as a function of the total energy $E$,
\begin{equation}
  \sigma_{vj,v'j'}(E) = \frac{\pi\hbar^2}{2\mu(E-E_{vj})(2j+1)}
    \sum_{J} (2J+1)\left|S_{vj,v'j'}^J(E)\right|^2
  \label{eq:xs}  
\end{equation}
and the state-to-state rate coefficient $k_{vj,v'j'}$ given as a function of temperature $T$,
\begin{equation}
  k_{vj,v'j'}(T) = \sqrt{\frac{8}{\pi\mu\left(k_B T\right)^3}}
    \int_{0}^{+\infty} dE_c \sigma_{vj,v'j'}(E_c)\,E_ce^{-E_c/k_BT} \,,
  \label{eq:rate}
\end{equation}
where $\hbar$ is the reduced Planck's constant, $k_B$ is Boltzman's constant, $\mu$ is the reduced mass of the reactants, $J$ is the total angular momentum of the H$_2$D$^+$ system, $S_{vj,v'j'}^J(E)$ are the elements of the scattering matrix, and $E_c=E-E_{vj}$ is the collision energy in the entrance channel $\mathrm{H}^+ + \mathrm{HD}(v,j)$.

The scattering matrix is calculated for a given total energy $E$ and total angular momentum $J$ using a fully Coriolis-coupled TIQM method based on the body-fixed democratic hyperspherical coordinates, and described in details in \citep{honvault2004}. At each hyperradius $\rho$, the scattering wave function is expanded on a set of appropriate hyperangular basis functions. The $\rho$-dependent coefficients are solutions of a set of coupled second-order differential equations, which are solved using the Johson-Manolopoulos log-derivative propagator \citep{manolopoulos1986}. The scattering wave function is computed up to the hyperradius $\rho_\mathrm{max} = 17.5\,a_0$, where the $S_{vj,v'j'}^J(E)$ matrix elements are extracted for many rovibrational levels of the reactant HD and the product H$_2$.
This method has been successfully applied to the isotopic variants $\mathrm{H}^+ + \mathrm{H}_2$ \citep{honvault2011, honvault2011b, rao2014, gonzalez-lezana2014, gonzalez-lezana2017}, $\mathrm{D}^+ + \mathrm{H}_2$ \citep{honvault2013, gonzalez-lezana2013, gonzalez-lezana2014, lara2015} and $\mathrm{H}^+ + \mathrm{D}_2$ \citep{gonzalez-lezana2009}. Here we use the VLABP PES calculated by \citet{velilla2008}, and which accurately describes the long-range interactions between H$^+$ and H$_2$.

\begin{table}
  \caption{Parameters of the intervals into which the energy grid is split: minimum $E_\mathrm{min}$, maximum $E_\mathrm{max}$ and increment $\Delta E$ in eV, as well as the maximum total angular momentum $J_\mathrm{max}$. \label{tab:ener}}
  \begin{tabular}{clllr}
   $(v,j)$ & $E_\mathrm{min}$ & $E_\mathrm{max}$ & $\Delta E$ & $J_\mathrm{max}$ \\
  \hline
   $(0,0)$ &            0.031  &             0.1  &      0.003 &               28 \\
   $(0,0)$ &            0.11   &             0.3  &      0.01  &               42 \\
   $(0,0)$ &            0.31   &             1    &      0.03  &               62 \\
   $(0,0)$ &            1.1    &             1.8  &      0.1   &               75 \\
   $(0,j\ge 2)$ &       0.0034 &             0.01 & $6\times 10^{-4}$ &        24 \\
   $(0,j\ge 2)$ &       0.011  &             0.03 &      0.001 &               28 \\
  \end{tabular}
\end{table}

Because the charge-transfer channel $\mathrm{H} + \mathrm{H}_2^+$ is located at an electronic energy of 1.83~eV above $\mathrm{H}^+ + \mathrm{H}_2$, we perform our scattering calculations up to 1.8~eV above the lowest entrance channel $\mathrm{H}^+ + \mathrm{HD}(v=0,j=0)$. Our energy grid, which contains about 80 points, is denser for low collision energies. The energy grid is split into intervals inside which we take the same maximum total angular momentum $J_\mathrm{max}$ (see Table \ref{tab:ener}). Regarding the ground rovibrational level of HD, the lowest collision energy that we take is 0.031~eV, for which the reaction is impossible, but we see the opening of the reactive channel at 0.037~eV. For each excited rotational level $J\ge 2$, we also consider collision energies down to 0.003~eV, a value below which our matching distance $\rho_{\mathrm{max}} = 17.5\,a_0$ does not allow for a satisfactory convergence.
On the other hand, to get a good convergence for an energy of 1.8~eV above the $\mathrm{H}^+ + \mathrm{HD}(v=0,j=0)$, we need to include 252 and 98 (open and closed) rovibrational levels for HD and H$_2$. In the case of HD, $j$ ranges from 0 to 29 for $v=0$, and $v$ ranges from 0 to 11 for $j=0$. As a consequence, we get 350, 2875 and 3892 coupled channels for $J=0$, 10 and 20, respectively.

\section{Results}
\label{sec:res}

In this section, we present cross sections and rate coefficients for selected initial (HD) and final (H$_2$) rovibrational levels. The collision energies are expressed in electron-volts (eV), cross sections in units of squared Bohr radius $a_0^2$, temperature in kelvin (K), and rate coefficients in \cms{.}

\subsection{HD in $(v=0,j=0)$}
\label{sub:j0}

\begin{figure}
  \includegraphics[width=\columnwidth]{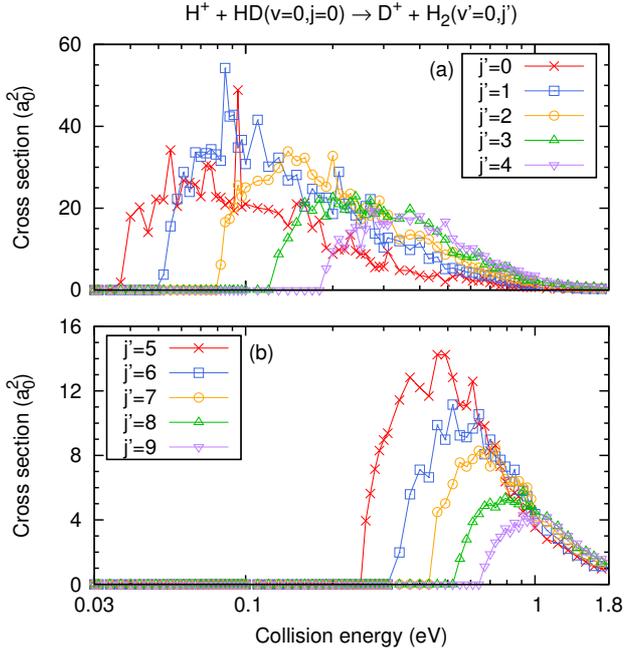}
  \caption{Cross sections as functions of the collision energy, for the reactions $\mathrm{H}^+ + \mathrm{HD}(v=0,j=0) \to \mathrm{D}^+ + \mathrm{H}_2(v'=0,j')$, with (a) $j'=0$ to 4, and (b) $j'=5$ to 9. \label{fig:xs_00}}
\end{figure}

Figure \ref{fig:xs_00} shows the cross sections of the reactions $(v=0,j=0) \to (v'=0,j')$ for $j'=0$ to 9, as functions of the collision energy. For all those endothermic reactions, the logarithmic scale enables us to see the opening of the reactive channel associated with each rotational level $j'$ of H$_2$. For energies slightly above that threshold, the cross sections abruptly increase, while they slowly decrease at high energies, down to values smaller than 2~$a_0^2$ at 1.8 eV. In addition, the curves present resonant peaks for energies lower than 0.5 eV. Those peaks are higher and more numerous for the low rotational levels of H$_2$: for instance, the highest peak is observed for $j'=1$ (54~$a_0^2$ at 0.085 eV). The curve with $j'=0$ also presents a narrow and high resonance (49~$a_0^2$ at 0.094 eV); but that curve is generally below the $j'=1$ one.

\begin{figure}
  \includegraphics[width=\columnwidth]{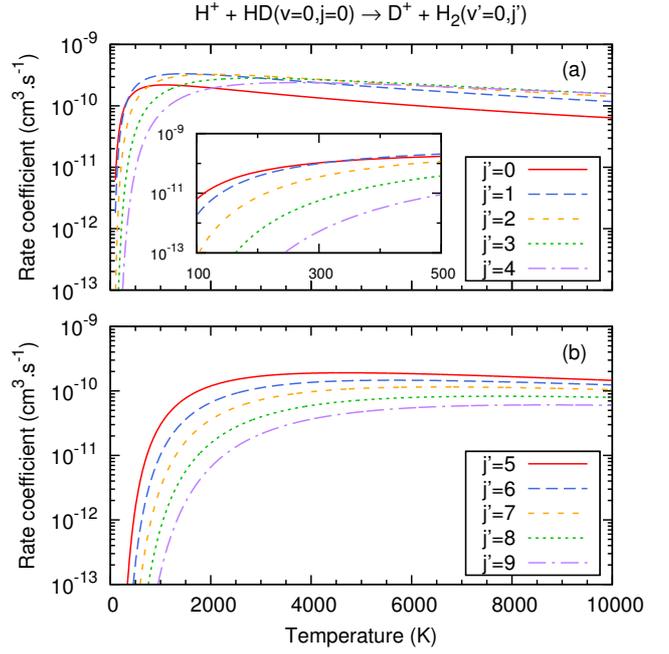}
  \caption{Rate coefficients as functions of the temperature, for the reactions $\mathrm{H}^+ + \mathrm{HD}(v=0,j=0) \to \mathrm{D}^+ + \mathrm{H}_2(v'=0,j')$, with (a) $j'=0$ to 4, and (b) $j'=5$ to 9. The inset in panel (a) shows a zoom between 100 and 500~K. \label{fig:rt_00}}
\end{figure}

Figure \ref{fig:rt_00} shows the rate constants as functions of the temperature between 100 and 10000~K, for the same reactions as Fig.~\ref{fig:xs_00}. All the curves evolve in a similar way with temperature, showing a fast increase, a maximum, and then a slow decrease. The low-temperature increase is faster with smaller $j'$, due to the lower energy thresholds observed on Figure \ref{fig:xs_00}. Moreover, as $j'$ increases, the high-temperature decrease in $k(T)$ is less pronounced; for example for $j'=9$, the bump is hardly visible since the maximum rate coefficient is equal to $6.06 \times 10^{-11}$~\cms{} (at 8700~K), while it is equal to $5.98 \times 10^{-11}$~\cms{} at 10000~K.

For levels $j'=5$ to 9, the rate coefficients are smaller on the whole range of temperatures of Fig.~\ref{fig:rt_00}(b) when $j'$ increases. But for lower values of $j'$, the hierarchy is less clear. Except for $T<300$~K, the reaction towards $j'=0$ is never the dominant one, as expected with cross sections (see Fig.~\ref{fig:xs_00}). The highest rate is obtained for $j'=1$ ($3.33 \times 10^{-10}$~\cms at 1450~K), but for higher temperatures, this rate becomes smaller than those for $j'=2$ to 5. At the temperature of 6000~K for instance, the dominant reactions are toward $j'=3$, 2 and 4.

It is important to stress that the high-temperature decrease of Fig.~ \ref{fig:rt_00} is not due to a bad convergence of the integral in Eq.~\eqref{eq:rate}. To check it, we have removed the last two cross sections (at 1.7 and 1.8 eV) in the calculation of rate coefficients. At 10000~K, we found a relative difference of 0.7, 1.2 and 8.7~\% for $j'=0$, 3 and 9, respectively. The convergence is less good for $j'=9$ because the cross section at 1.8 eV, equal to 1.46~$a_0^2$ represents one third of its maximum value (4.30~$a_0^2$ at 0.91 eV); its removal has thus a sizeable influence on the integral \eqref{eq:rate}.

\begin{figure}
  \includegraphics[width=\columnwidth]{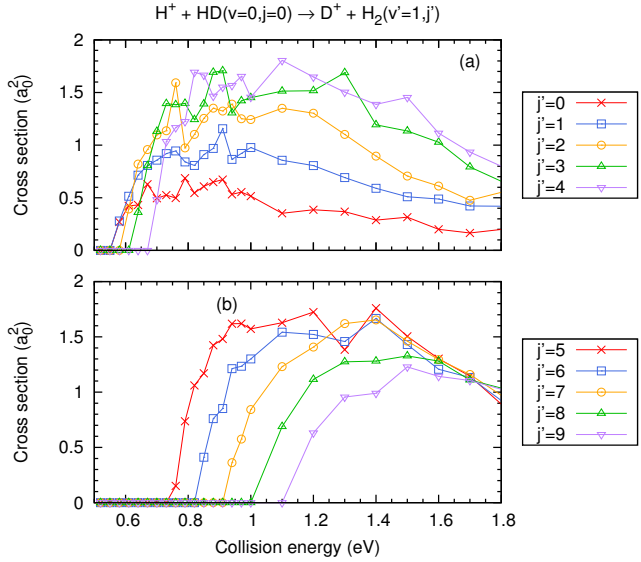}
  \caption{Cross sections as functions of the collision energy, for the reactions $\mathrm{H}^+ + \mathrm{HD}(v=0,j=0) \to \mathrm{D}^+ + \mathrm{H}_2(v'=1,j')$, with (a) $j'=0$ to 4, and (b) $j'=5$ to 9. \label{fig:xs_01}}
\end{figure}

\begin{figure}
  \includegraphics[width=\columnwidth]{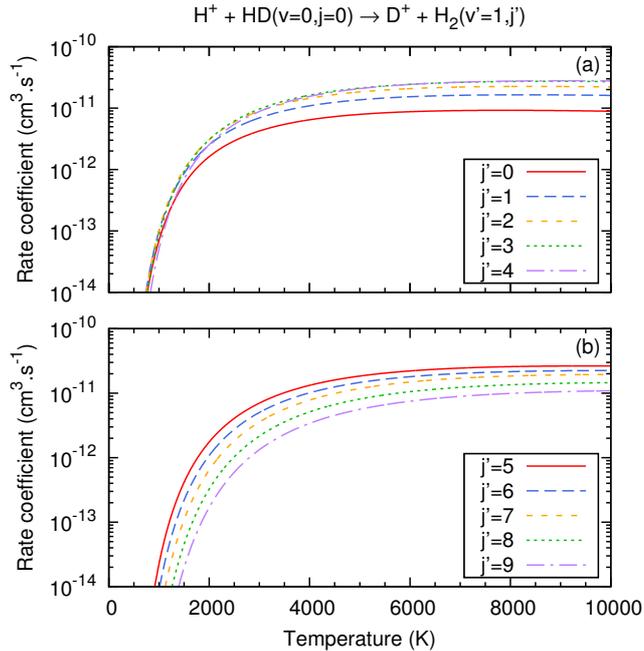}
  \caption{Rate coefficients as functions of the temperature, for the reactions $\mathrm{H}^+ + \mathrm{HD}(v=0,j=0) \to \mathrm{D}^+ + \mathrm{H}_2(v'=1,j')$, with (a) $j'=0$ to 4, and (b) $j'=5$ to 9. \label{fig:rt_01}}
\end{figure}

Now we consider the product H$_2$ in its first excited vibrational level $v'=1$. The cross sections are plotted in Fig.~\ref{fig:xs_01}, in an energy range between 0.5 and 1.8 eV and in linear scale. In that range, the energy resolution does not allow for distinguishing the thresholds of the reactions $(0,0) \to (1,0)$ and $(0,0) \to (1,1)$; but it is sufficient in order to understand the overall evolution of the cross section, which is relevant for the calculation of rate coefficients. Globally, the cross sections for $v'=1$ are much smaller than for $v'=0$ (see Fig.~\ref{fig:xs_00}. This trend is also visible for the rate coefficients, plotted in Fig.~\ref{fig:rt_01}. For example at 1000~K, the rate coefficients for $v'=1$ are on the order of $10^{-13}$~\cms or below, that is three orders of magnitude lower than those for $v'=0$. This gap decreases with temperature, and at 10000~K, there is approximately a one-order-of-magnitude difference between $v'=0$ and $v'=1$. As for the $j'$-dependence of the rates, at 6000~K, the dominant reactions involve intermediate $j'$-values (3, 4 then 5), while $j'=0$ has the second smallest rate.

We have also checked the convergence of the rate coefficients given in fig.~\ref{fig:rt_01} by removing the last two cross sections of Fig.~\ref{fig:xs_01}. This reduces the rate coefficients by 5.9~\% and 42~\% for $j'=0$ and 9 respectively. The latter value is thus less accurate, but one should keep in mind that corresponding reaction is by far not the dominant one driving to the destruction of HD($v=0,j=0)$.

\begin{figure}
  \includegraphics[width=\columnwidth]{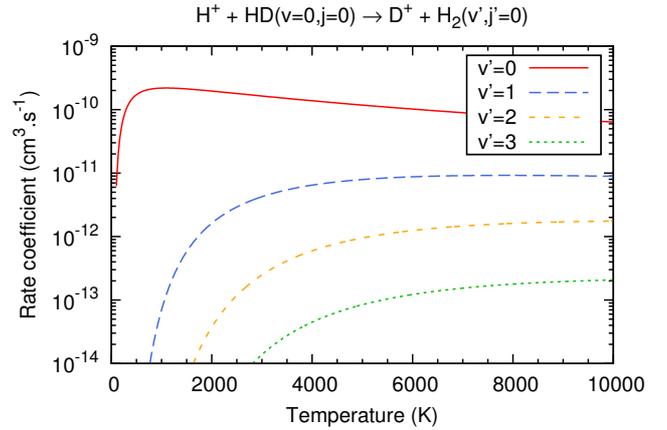}
  \caption{Rate coefficients as functions of the temperature, for the reactions $\mathrm{H}^+ + \mathrm{HD}(v=0,j=0) \to \mathrm{D}^+ + \mathrm{H}_2(v',j'=0)$, with $v'=0$ to 3. \label{fig:rt_02}}
\end{figure}

When $v'=2$ and 3, the rate coefficients are even smaller, as shown in fig.~\ref{fig:rt_02}: at 6000~K, the rates lose one order of magnitude when $v'$ increases by one unity. Regarding the convergence of equation \eqref{eq:rate}, for $v'=3$, there are only five non-zero cross sections, from 1.4 to 1.8~eV; removing the last value decreases the rate coefficients by 38~\%.

\subsection{HD in excited rotational levels $(v=0,j)$}
\label{sub:j1}

\begin{figure}
  \includegraphics[width=\columnwidth]{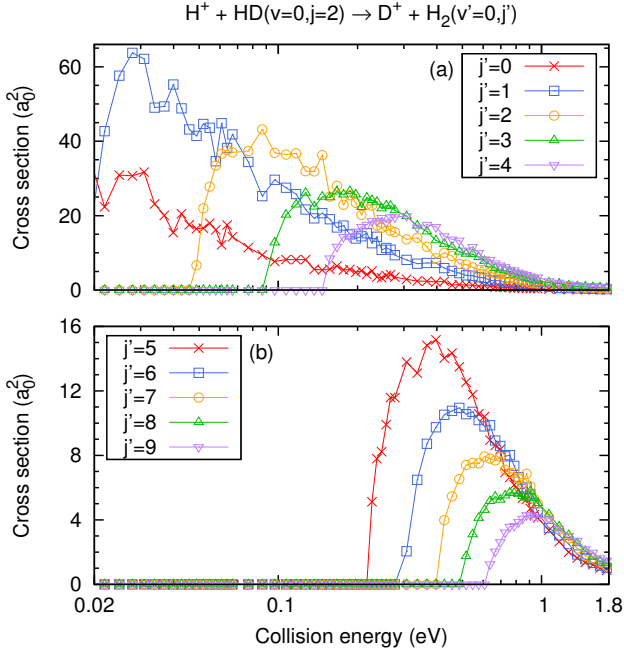}
  \caption{Cross sections as functions of the collision energy, for the reactions $\mathrm{H}^+ + \mathrm{HD}(v=0,j=2) \to \mathrm{D}^+ + \mathrm{H}_2(v'=0,j')$, with (a) $j'=0$ to 4, and (b) $j'=5$ to 9. \label{fig:xs_20}}
\end{figure}

\begin{figure}
  \includegraphics[width=\columnwidth]{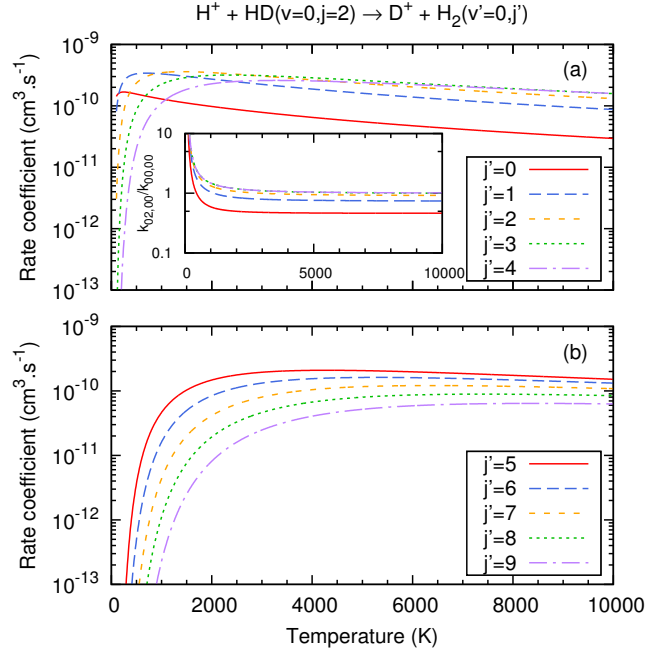}
  \caption{Rate coefficients as functions of the temperature, for the reactions $\mathrm{H}^+ + \mathrm{HD}(v=0,j=2) \to \mathrm{D}^+ + \mathrm{H}_2(v'=0,j')$, with (a) $j'=0$ to 4, and (b) $j'=5$ to 9. The inset of panel (a) shows the ratio between rate coefficients $k_{02,0j'}(T) / k_{00,0j'}(T)$. \label{fig:rt_20}}
\end{figure}

For the first rotationally excited level $j=1$, the results, very similar to the previous ones, are given in the supplementary material. In particular, at 6000~K, the largest reactions rate involve $j'=3$, 4 then 2. The case $j=2$ is a little particular, since the reaction $\mathrm{H}^+ + \mathrm{HD}(v=0,j=2) \to \mathrm{D}^+ + \mathrm{H}_2(v'=0,j'=0)$ is endothermic by only 3.4~meV (neglecting the difference in ionisation energies between H and D). So for a given $j'$-value, the energy thresholds for $(0,2) \to (0,j')$, see Fig.~\ref{fig:xs_20}, are lower than for $(0,0) \to (0,j')$, see Fig.~\ref{fig:xs_00}, and the maximum of the curve has a smaller energy and a larger cross section. The consequence on the rate coefficients can be seen in Fig.~\ref{fig:rt_20}: they are all larger for the reaction $(0,2) \to (0,j')$ at low temperatures; but as $T$ increases, the ratio $k_{02,0j'}(T) / k_{00,0j'}(T)$ decreases, and becomes smaller than unity for the lowest $j'$-values (see insets of Fig.~\ref{fig:rt_20}).

\begin{figure}
  \includegraphics[width=\columnwidth]{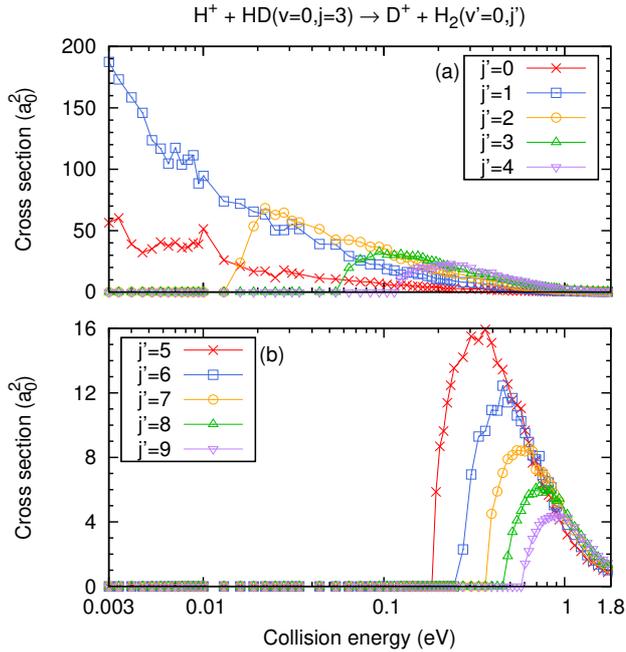}
  \caption{Cross sections as functions of the collision energy, for the reactions $\mathrm{H}^+ + \mathrm{HD}(v=0,j=3) \to \mathrm{D}^+ + \mathrm{H}_2(v'=0,j')$, with (a) $j'=0$ to 4, and (b) $j'=5$ to 9. \label{fig:xs_30}}
\end{figure}

\begin{figure}
  \includegraphics[width=\columnwidth]{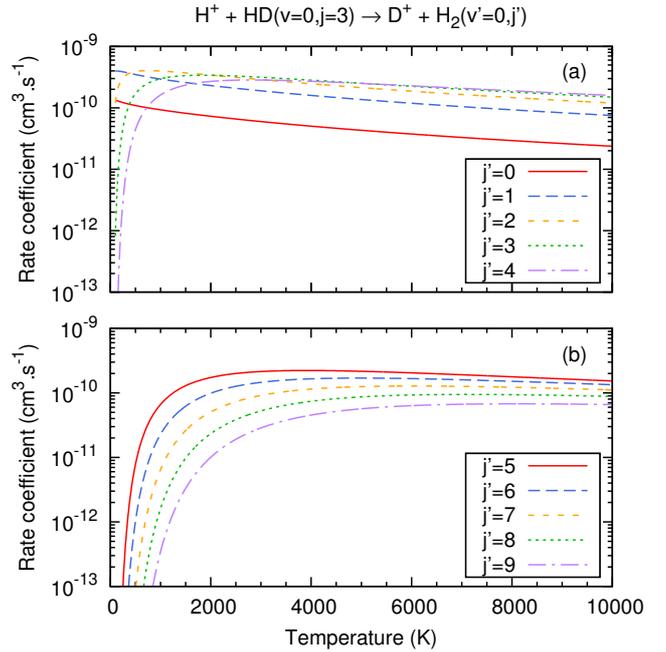}
  \caption{Rate coefficients as functions of the temperature, for the reactions $\mathrm{H}^+ + \mathrm{HD}(v=0,j=3) \to \mathrm{D}^+ + \mathrm{H}_2(v'=0,j')$, with (a) $j'=0$ to 4, and (b) $j'=5$ to 9. \label{fig:rt_30}}
\end{figure}

The case $j=3$ is particularly interesting, since the reactions $(0,3) \to (0,0)$ and $(0,3) \to (0,1)$ are exothermic by 0.029 and 0.015~eV. Figure \ref{fig:xs_30} shows the corresponding cross sections down to an energy of 3~meV, a value below which our value of $\rho_{\mathrm{max}} = 17.5~a_0$ does not allow for a full convergence of the cross section. The reaction $(0,3) \to (0,1)$ strongly dominates over $(0,3) \to (0,0)$, for instance by a factor of 3 at 3~meV. This predominance is also visible on rate coefficients, see Fig.~\ref{fig:rt_30}. On the whole range of temperature, $k_{03,01}$ is larger than $k_{03,00}$ by a factor of $\simeq 3$; but is exceeded by $k_{03,02}$ for $T\ge 400$~K and then by other rates. At large temperatures, the results are similar to other $j$-values: the rates slowly decrease, and at 6000~K, the dominant reactions are towards $j'=4$, 3 and 5.

In order to check the convergence at low temperature, we have compared the previous rates with those calculated by removing the first two cross sections (at 3 et 3.4~meV) in equation \eqref{eq:rate}. At 100~K, the rates $k_{03,00}$ and $k_{03,01}$ decrease by 6.5~\% and 6.9~\% respectively. At large temperature, the convergence is very similar to the one of $k_{00,00}$. 

In the case $j=4$, there are three exothermic reactions, for $j'=0$, 1 and 2. The cross sections and rate coefficients, which are given in the supplementary material, look like Figs.~\ref{fig:xs_30} and \ref{fig:rt_30}. At low energies, the cross section $\sigma_{04,02}$ is the largest ($169~a_0^2$ at 3~meV), followed by $\sigma_{04,01}$ ($128~a_0^2$) and $\sigma_{04,00}$ ($36~a_0^2$). This trend shows up on low-temperature rates, while the high-temperature rates are similar to the other $j$-values: at 6000~K, the largest rates are towards $j'=4$, 3 and 5.

\subsection{HD in thermodynamic equilibrium}
\label{sub:therm-eq}

In order to compare our results with literature, we have computed the thermal rate coefficient $k(T)$ of the title reaction, assuming that the rovibrational levels of HD are populated according to a Maxwell-Boltzman distribution. Since we deal with rather low temperatures in this paragraph, we can assume that only the five lowest rotational levels $(v=0,j\in[0;4])$ are significantly populated. For instance at 295~K, the population in $(0,0)$ is equal to 20.1~\%, that of $(0,1)$ is 39.1~\%, while those of $(0,4)$ and $(0,5)$ are 2.4~\% and 0.4~\% respectively. Moreover, as shown in Fig.~\ref{fig:rt_20}, the reactions producing vibrationally excited H$_2$ can safely be ignored. Therefore, our thermal rate coefficient is given by
\begin{equation}
  k(T) = \frac{1}{Z_\mathrm{rot}} \sum_{j=0}^4 (2j+1) 
    e^{-E_{vj}/k_BT} \sum_{j'=0}^9 k_{0j,0j'}(T)
  \label{eq:therm}
\end{equation}
where $Z_\mathrm{rot} = \sum_{j=0}^4 (2j+1) e^{-E_{vj}/k_BT}$ is the rotational partition function restricted to the five lowest rotational levels of HD.

By doing so, we find a thermal rate coefficient that can be very well fitted with an Arrhenius formula $A\times\exp(-B/T)$, between 100 and 400~K, with $A=1.84 \times 10^{-9}$~\cms{} and $B=416$~K. More specifically, we obtain $k(T=205~K) = 2.4$ and $k(T=295~K) = 4.5 \times 10^{-10}$~\cms{,} which are approximately twice as large as the measurements of Refs.~\citep{henchman1981, smith1982}, and also significantly larger the rates calculated by \citet{jambrina2012}. In view of these discrepancies, we have checked the quality of our scattering matrices, by using them to calculate cross sections and rate coefficients of the reverse reaction $\mathrm{D}^+ + \mathrm{H}_2 \to \mathrm{H}^+ + \mathrm{HD}$: our results are in very good agreement with former calculations \citep{honvault2013}, themselves in very good agreement with experiments for collision energies larger than 0.01~eV \citep{gerlich1992}.

The discrepancies obtained with the TIQM calculations of \citet{jambrina2012} may come from the PESs employed. Indeed, \citet{rao2014} have shown that the cross sections and rate coefficients of the $\mathrm{H}^+ + \mathrm{H}_2$ reaction are always significantly larger with the VLABP PES than with the KBNNPES by \citet{kamisaka2002}. But as shown in \citep{gonzalez-lezana2009} on the $\mathrm{H}^+ + \mathrm{D}_2$ reaction, there is not such a heavy trend between the KBNN and ARTSM PESs. Extending those results to the title reaction thus suggests that the ARTSM and VLABP PESs are likely to give significantly different cross sections and rate coefficients. Because the calculations of \citet{jambrina2012} were performed with the ARTSM PES, this may explain why our rate coefficients overcome those of \citet{jambrina2012}. In any case, additional measurements, with more collision energies or temperatures, would be necessary.

\section{Conclusion}
\label{sec:ccl}

We have computed cross sections and rate coefficients of the state-to-state $\mathrm{H}^+ + \mathrm{HD}(v,j) \to \mathrm{D}^+ + \mathrm{H}_2(v',j')$ reaction, where HD is in the vibrational ground level. The rate coefficients are computed for temperatures from 100 to 10000~K. They are significantly smaller for vibrationally excited H$_2$ molecules, approximately losing one order of magnitude when $v'$ increases by one. For temperature below 4000~K, rate coefficients also strongly depend on the initial rotational level, as a function of which the reaction is endothermic or exothermic. Investing state-to-state $\mathrm{H}^+ + \mathrm{HD} \to \mathrm{D}^+ + \mathrm{H}_2$ reactions is therefore crucial to model the chemical evolution of the primordial gas. In a future work, we will consider HD in excited vibrational levels, for which the reaction is exothermic for all rotational levels.

In order to compare our results with the few literature ones, we have also calculated thermal rate coefficients, assuming the HD rovibrational levels in thermodynamic equilibrium. We find rate coefficients roughly twice as large as published values. Regarding previous theoretical results, we possibly attribute those discrepancies to differences in the asymptotic region of the potential-energy surfaces used for the calculations. In any case, more experimental results are necessary, since thermal rate coefficients have only been measured for two temperatures.

\section*{Acknowledgements}

M.L. acknowledges the financial support of {}``R{\'e}gion Bourgogne Franche Comt{\'e}'' under the projet 2018Y.07063 {}``Th{\'e}CUP''. This work was supported by the Programme National {}``Physique et Chimie du Milieu Interstellaire'' (PCMI) of CNRS/INSU with INC/INP co-funded by CEA and CNES.




\begin{thebibliography}{}
\makeatletter
\relax
\def\mn@urlcharsother{\let\do\@makeother \do\$\do\&\do\#\do\^\do\_\do\%\do\~}
\def\mn@doi{\begingroup\mn@urlcharsother \@ifnextchar [ {\mn@doi@}
  {\mn@doi@[]}}
\def\mn@doi@[#1]#2{\def\@tempa{#1}\ifx\@tempa\@empty \href
  {http://dx.doi.org/#2} {doi:#2}\else \href {http://dx.doi.org/#2} {#1}\fi
  \endgroup}
\def\mn@eprint#1#2{\mn@eprint@#1:#2::\@nil}
\def\mn@eprint@arXiv#1{\href {http://arxiv.org/abs/#1} {{\tt arXiv:#1}}}
\def\mn@eprint@dblp#1{\href {http://dblp.uni-trier.de/rec/bibtex/#1.xml}
  {dblp:#1}}
\def\mn@eprint@#1:#2:#3:#4\@nil{\def\@tempa {#1}\def\@tempb {#2}\def\@tempc
  {#3}\ifx \@tempc \@empty \let \@tempc \@tempb \let \@tempb \@tempa \fi \ifx
  \@tempb \@empty \def\@tempb {arXiv}\fi \@ifundefined
  {mn@eprint@\@tempb}{\@tempb:\@tempc}{\expandafter \expandafter \csname
  mn@eprint@\@tempb\endcsname \expandafter{\@tempc}}}

\bibitem[\protect\citeauthoryear{Aguado, Roncero, Tablero, Sanz  \&
  Paniagua}{Aguado et~al.}{2000}]{aguado2000}
Aguado A.,  Roncero O.,  Tablero C.,  Sanz C.,   Paniagua M.,  2000, J. Chem.
  Phys., 112, 1240

\bibitem[\protect\citeauthoryear{Bhowmick, Bossion, Scribano  \&
  Suleimanov}{Bhowmick et~al.}{2018}]{bhowmick2018}
Bhowmick S.,  Bossion D.,  Scribano Y.,   Suleimanov Y.,  2018, Phys. Chem.
  Chem. Phys., 20, 26752

\bibitem[\protect\citeauthoryear{Bovino \& Galli}{Bovino \&
  Galli}{2018}]{bovino2018}
Bovino S.,  Galli D.,  2018, arXiv preprint arXiv:1807.05939

\bibitem[\protect\citeauthoryear{Coppola, Longo, Capitelli, Palla  \&
  Galli}{Coppola et~al.}{2011}]{coppola2011}
Coppola C.,  Longo S.,  Capitelli M.,  Palla F.,   Galli D.,  2011, Astrophys.
  J. Suppl. Ser., 193, 7

\bibitem[\protect\citeauthoryear{Coppola, Galli, Palla, Longo  \&
  Chluba}{Coppola et~al.}{2013}]{coppola2013}
Coppola C.,  Galli D.,  Palla F.,  Longo S.,   Chluba J.,  2013, Mon. Not. R.
  Astron. Soc., 434, 114

\bibitem[\protect\citeauthoryear{Coppola, Mizzi, Bruno, Esposito, Galli, Palla
  \& Longo}{Coppola et~al.}{2016}]{coppola2016}
Coppola C.,  Mizzi G.,  Bruno D.,  Esposito F.,  Galli D.,  Palla F.,   Longo
  S.,  2016, Mon. Not. R. Astron. Soc., 457, 3732

\bibitem[\protect\citeauthoryear{Desrousseaux, Coppola, Kazandjian  \&
  Lique}{Desrousseaux et~al.}{2018}]{desrousseaux2018}
Desrousseaux B.,  Coppola C.,  Kazandjian M.,   Lique F.,  2018, J. Phys. Chem.
  A, 122, 8390

\bibitem[\protect\citeauthoryear{Ely, Coppola  \& Lique}{Ely
  et~al.}{2016}]{ely2016}
Ely S.,  Coppola C.,   Lique F.,  2016, Mon. Not. R. Astron. Soc., 466, 2175

\bibitem[\protect\citeauthoryear{Fehsenfeld, Albritton, Bush, Fournier, Govers
  \& Fournier}{Fehsenfeld et~al.}{1974}]{fehsenfeld1974}
Fehsenfeld F.,  Albritton D.,  Bush Y.,  Fournier P.,  Govers T.,   Fournier
  J.,  1974, J. Chem. Phys., 61, 2150

\bibitem[\protect\citeauthoryear{Flower \& Roueff}{Flower \&
  Roueff}{1999}]{flower1999}
Flower D.,  Roueff E.,  1999, Mon. Not. R. Astron. Soc., 309, 833

\bibitem[\protect\citeauthoryear{Galli \& Palla}{Galli \&
  Palla}{2002}]{galli2002}
Galli D.,  Palla F.,  2002, Planetary and Space Science, 50, 1197

\bibitem[\protect\citeauthoryear{Galli \& Palla}{Galli \&
  Palla}{2013}]{galli2013}
Galli D.,  Palla F.,  2013, Ann. Rev. Astron. Astrophys., 51, 163

\bibitem[\protect\citeauthoryear{Gerlich}{Gerlich}{1992}]{gerlich1992}
Gerlich D.,  1992, Adv. Chem. Phys., 82, 1

\bibitem[\protect\citeauthoryear{Gerlich \& Schlemmer}{Gerlich \&
  Schlemmer}{2002}]{gerlich2002}
Gerlich D.,  Schlemmer S.,  2002, Planetary and Space Science, 50, 1287

\bibitem[\protect\citeauthoryear{Ghosh, Sahoo, Adhikari, Sharma  \&
  Varandas}{Ghosh et~al.}{2015}]{ghosh2015}
Ghosh S.,  Sahoo T.,  Adhikari S.,  Sharma R.,   Varandas A.,  2015, J. Phys.
  Chem. A, 119, 12392

\bibitem[\protect\citeauthoryear{Ghosh, Mukherjee, Mukherjee, Mandal, Sharma,
  Chaudhury  \& Adhikari}{Ghosh et~al.}{2017}]{ghosh2017}
Ghosh S.,  Mukherjee S.,  Mukherjee B.,  Mandal S.,  Sharma R.,  Chaudhury P.,
   Adhikari S.,  2017, J. Chem. Phys., 147, 074105

\bibitem[\protect\citeauthoryear{Gonz{\'a}lez-Lezana \&
  Honvault}{Gonz{\'a}lez-Lezana \& Honvault}{2017}]{gonzalez-lezana2017}
Gonz{\'a}lez-Lezana T.,  Honvault P.,  2017, Mon. Not. R. Astron. Soc., 467,
  1294

\bibitem[\protect\citeauthoryear{Gonz{\'a}lez-Lezana, Honvault, Jambrina, Aoiz
  \& Launay}{Gonz{\'a}lez-Lezana et~al.}{2009}]{gonzalez-lezana2009}
Gonz{\'a}lez-Lezana T.,  Honvault P.,  Jambrina P.,  Aoiz F.,   Launay J.-M.,
  2009, J. Chem. Phys., 131, 044315

\bibitem[\protect\citeauthoryear{Gonz{\'a}lez-Lezana, Honvault  \&
  Scribano}{Gonz{\'a}lez-Lezana et~al.}{2013}]{gonzalez-lezana2013}
Gonz{\'a}lez-Lezana T.,  Honvault P.,   Scribano Y.,  2013, J. Chem. Phys.,
  139, 054301

\bibitem[\protect\citeauthoryear{Gonz{\'a}lez-Lezana, Scribano  \&
  Honvault}{Gonz{\'a}lez-Lezana et~al.}{2014}]{gonzalez-lezana2014}
Gonz{\'a}lez-Lezana T.,  Scribano Y.,   Honvault P.,  2014, J. Phys. Chem. A,
  118, 6416

\bibitem[\protect\citeauthoryear{Henchman, Adams  \& Smith}{Henchman
  et~al.}{1981}]{henchman1981}
Henchman M.,  Adams N.,   Smith D.,  1981, J. Chem. Phys., 75, 1201

\bibitem[\protect\citeauthoryear{Honvault \& Launay}{Honvault \&
  Launay}{2004}]{honvault2004}
Honvault P.,  Launay J.-M.,  2004, in , Theory of Chemical Reaction Dynamics.
Springer, pp 187--215

\bibitem[\protect\citeauthoryear{Honvault \& Scribano}{Honvault \&
  Scribano}{2013}]{honvault2013}
Honvault P.,  Scribano Y.,  2013, J. Phys. Chem. A, 117, 9778

\bibitem[\protect\citeauthoryear{Honvault, Jorfi, Gonz{\'a}lez-Lezana, Faure
  \& Pagani}{Honvault et~al.}{2011a}]{honvault2011b}
Honvault P.,  Jorfi M.,  Gonz{\'a}lez-Lezana T.,  Faure A.,   Pagani L.,
  2011a, Phys. Chem. Chem. Phys., 13, 19089

\bibitem[\protect\citeauthoryear{Honvault, Jorfi, Gonz{\'a}lez-Lezana, Faure
  \& Pagani}{Honvault et~al.}{2011b}]{honvault2011}
Honvault P.,  Jorfi M.,  Gonz{\'a}lez-Lezana T.,  Faure A.,   Pagani L.,
  2011b, Phys. Rev. Lett., 107, 023201

\bibitem[\protect\citeauthoryear{Jambrina, Aoiz, Eyles, Herrero  \&
  S{\'a}ez~R{\'a}banos}{Jambrina et~al.}{2009}]{jambrina2009}
Jambrina P.,  Aoiz F.,  Eyles C.,  Herrero V.,   S{\'a}ez~R{\'a}banos V.,
  2009, J. Chem. Phys., 130, 184303

\bibitem[\protect\citeauthoryear{Jambrina, Alvari{\~n}o, Gerlich, Hankel,
  Herrero, S{\'a}ez-R{\'a}banos  \& Aoiz}{Jambrina et~al.}{2012}]{jambrina2012}
Jambrina P.,  Alvari{\~n}o J.,  Gerlich D.,  Hankel M.,  Herrero V.,
  S{\'a}ez-R{\'a}banos V.,   Aoiz F.,  2012, Phys. Chem. Chem. Phys., 14, 3346

\bibitem[\protect\citeauthoryear{Kamisaka, Bian, Nobusada  \&
  Nakamura}{Kamisaka et~al.}{2002}]{kamisaka2002}
Kamisaka H.,  Bian W.,  Nobusada K.,   Nakamura H.,  2002, J. Chem. Phys., 116,
  654

\bibitem[\protect\citeauthoryear{Kreckel, Bruhns, {\v{C}}{\'\i}{\v{z}}ek,
  Glover, Miller, Urbain  \& Savin}{Kreckel et~al.}{2010}]{kreckel2010}
Kreckel H.,  Bruhns H.,  {\v{C}}{\'\i}{\v{z}}ek M.,  Glover S.,  Miller K.,
  Urbain X.,   Savin D.,  2010, Science, 329, 69

\bibitem[\protect\citeauthoryear{Lara, Jambrina, Aoiz  \& Launay}{Lara
  et~al.}{2015}]{lara2015}
Lara M.,  Jambrina P.,  Aoiz F.,   Launay J.-M.,  2015, J. Chem. Phys., 143,
  204305

\bibitem[\protect\citeauthoryear{Manolopoulos}{Manolopoulos}{1986}]{manolopoulos1986}
Manolopoulos D.,  1986, J. Chem. Phys., 85, 6425

\bibitem[\protect\citeauthoryear{Millar, Bennett  \& Herbst}{Millar
  et~al.}{1989}]{millar1989}
Millar T.,  Bennett A.,   Herbst E.,  1989, Astrophys. J., 340, 906

\bibitem[\protect\citeauthoryear{Rao, Mahapatra  \& Honvault}{Rao
  et~al.}{2014}]{rao2014}
Rao T.,  Mahapatra S.,   Honvault P.,  2014, J. Chem. Phys., 141, 064306

\bibitem[\protect\citeauthoryear{Ripamonti}{Ripamonti}{2007}]{ripamonti2007}
Ripamonti E.,  2007, Mon. Not. R. Astron. Soc., 376, 709

\bibitem[\protect\citeauthoryear{Sahoo, Ghosh, Adhikari, Sharma  \&
  Varandas}{Sahoo et~al.}{2014}]{sahoo2014}
Sahoo T.,  Ghosh S.,  Adhikari S.,  Sharma R.,   Varandas A.,  2014, J. Phys.
  Chem. A, 118, 4837

\bibitem[\protect\citeauthoryear{Sahoo, Ghosh, Adhikari, Sharma  \&
  Varandas}{Sahoo et~al.}{2015}]{sahoo2015}
Sahoo T.,  Ghosh S.,  Adhikari S.,  Sharma R.,   Varandas A.,  2015, J. Chem.
  Phys., 142, 024304

\bibitem[\protect\citeauthoryear{Smith, Adams  \& Alge}{Smith
  et~al.}{1982}]{smith1982}
Smith D.,  Adams N.,   Alge E.,  1982, Astrophys. J., 263, 123

\bibitem[\protect\citeauthoryear{Velilla, Lepetit, Aguado, Beswick  \&
  Paniagua}{Velilla et~al.}{2008}]{velilla2008}
Velilla L.,  Lepetit B.,  Aguado A.,  Beswick J.,   Paniagua M.,  2008, J.
  Chem. Phys., 129, 084307

\bibitem[\protect\citeauthoryear{Viegas, Alijah  \& Varandas}{Viegas
  et~al.}{2007}]{viegas2007}
Viegas L.,  Alijah A.,   Varandas A.,  2007, J. Chem. Phys., 126, 074309

\bibitem[\protect\citeauthoryear{Yu, Su, Dai, Yuan  \& Yang}{Yu
  et~al.}{2014}]{yu2014}
Yu S.,  Su S.,  Dai D.,  Yuan K.,   Yang X.,  2014, J. Chem. Phys., 140, 034310

\makeatother
\end{thebibliography}



\bsp	
\label{lastpage}
\end{document}